\documentclass[%
 reprint,
superscriptaddress,
 amsmath,amssymb,
 aps,
 prapplied,
]{revtex4-2}

\usepackage{graphicx}
\usepackage{dcolumn}
\usepackage{bm}
\usepackage{xcolor}
\usepackage{soul}
\usepackage[normalem]{ulem}
\usepackage{amsmath}

\begin{document}

\title{Time resolved eye diagrams to exploit hidden high energy branches in a nonlinear wideband vibration energy harvester}

\author{Kankana Paul}
\email{kankana.paul@tyndall.ie}
\affiliation{Micropower-Nanomagnetics group, Micro-Nano-Systems Center, Tyndall National Institute, Cork, Ireland}

\author{Saibal Roy}
\email{saibal.roy@tyndall.ie}
\affiliation{Micropower-Nanomagnetics group, Micro-Nano-Systems Center, Tyndall National Institute, Cork, Ireland}
\affiliation{Department of Physics, University College Cork, Cork, Ireland}

\author{Andreas Amann}
\email{a.amann@ucc.ie}
\affiliation{School of Mathematical Science, University College Cork, Cork, Ireland}

\begin{abstract}
A wideband vibration energy harvester with multiple nonlinear forces is investigated. The nonlinearities are due to repulsive magnets and hardening springs, which gives rise to  multistabilities between a number of energy branches. Not all branches are accessible by a simple up or down sweep of the driving frequency and in particular the highest energy branch is often \textit{hidden}, requiring a suitable frequency schedule to be accessed. Detailed theoretical understanding of the energy branch structure along with robust experimental methods are essential for characterizing each of the energy branches to enhance the energy output from such vibration energy harvesting system. We introduce a graphical representation in the form of \textit{eye diagrams} based on time-resolved measurements of acceleration and output voltage to study the dynamical features of the different branches. This generic approach allows us to optimize the design, which results in 1.3mW of power generated at 1g over 44Hz frequency bandwidth while maintaining a small footprint of $1.23 cm^3$. The energy conversion ratio of the energy harvester at 120Hz drive frequency is 0.52 for the high energy branch.
\end{abstract}

\maketitle

\section{Introduction}

In this epoch of Internet of Things (IoT), the lack of a sustainable power source significantly impedes the pervasive deployment of autonomous sensors nodes. To address this cardinal issue, Vibration Energy Harvesters (VEHs) have emerged as a promising renewable energy source \cite{gao2016design,yang2019gullwing} due to the abundance of vibrations in the domestic and industrial environment. However, the characteristically narrow frequency bandwidth, and hence the poor off-resonance performance of traditional linear VEHs \cite{cammarano2014bandwidth} makes them unsuitable for harnessing substantial mechanical energy from ambient vibrations, which is spread over a broad spectrum of frequency.  The challenge is therefore to design a VEH with large energy output over a wide frequency range. 

A wider operable bandwidth is obtainable by using a VEH with a nonlinear restoring force \cite{PhysRevLett.102.080601,roy2015nonlinear} and in the past VEHs possessing monostable \cite{marinkovic2009smart,fan2018monostable, mallick2014nonlinear}, bistable \cite{yan2020scavenging,podder2015bistable, roy2021broadband}, tristable \cite{zhou2014broadband}, quadstable \cite{zhou2018harvesting}, and polystable\cite{deng2019poly} potential energy functions have been studied experimentally. From a theoretical point of view, VEHs can be modelled as driven nonlinear oscillators \cite{PhysRevE.83.066205}, which also appear in many other fields, including optics, photonics \cite{pickup2018optical,mueller2009optomechanical, PhysRevA.100.013813}, biomechanics \cite{harne2015dipteran}, and electronics \cite{chua1984nonlinear}. It is well known that, even in simple periodically driven systems the presence of nonlinearity can give rise to complex dynamical features, including multistability, dynamic symmetry breaking and frequency locking \cite{PhysRevApplied.13.014049,wieczorek2005dynamical,parlitz1985superstructure,PhysRevE.57.1563}.  

In the context of VEHs, the phenomenon of multistability translates into the presence of multiple \emph{energy branches} which coexist for a given set of driving parameters.  For example, in the classical case of a hardening nonlinearity \cite{PhysRevApplied.13.014011} as in the Duffing oscillator \cite{parlitz1985superstructure,PhysRevApplied.7.064002}, high-energy and low-energy branches coexist. The selection of the dynamical state depends on the initial conditions and the frequency schedule of the drive. Additionally, fully isolated resonances with large amplitudes are also possible \cite{marchionne2018synchronisation}.  

From an application point of view, it is desirable to maintain the system in the branch with the highest energy output, and various mechanisms for achieving and sustaining these high energy branches have been devised in the past \cite{mallick2016surfing,wang2019attaining, udani2017sustaining}. Thus a route to further increase the energy output and frequency bandwidth using more sophisticated nonlinearities appears possible. However, the increased complexity of the resulting energy branch structure requires detailed theoretical understanding and powerful experimental methods to characterize different branches.

The graphical representation of the dynamics of linear and nonlinear oscillators is a powerful tool for the estimation of energy generation \cite{kwak2021optimal}, energy transfer as well as the comparison of the performance with an ideal oscillator \cite{hosseinloo2015fundamental}. Particularly, the area enclosed in the force-displacement plane is useful for the investigation of the involved damping mechanism and the energy dissipated through the oscillator \cite{li2012electromagnetic,lv2015dielectric}. However, the potential of this method for characterizing complex nonlinear systems and the associated energy branches, which could lead to a more efficient energy harvesting system, is still unexplored.

In this work, we present a wideband vibration energy harvester that combines nonlinear forces arising from the spring-hardening and from repulsive magnetic interactions in a single device. We investigated the complex energy branch structure in this case. It was found that the highest-energy branch may be \emph{hidden} in the sense that a particular frequency schedule is required to reach it. Using our knowledge of the branch structure, we achieved this high energy branch even at a low level of excitation. To characterize the various energy branches experimentally, we took time-resolved measurements of acceleration and voltage outputs. This allows us to plot \textit{eye diagrams} in a force-displacement plane, where the enclosed areas (\textit{eyes}) represent the energy transacted within one period of the external drive. The different \textit{eye} shapes bear useful information about the nonlinearities involved and allow us to efficiently characterise the various energy branches experimentally. The visual representation using \textit{eye-diagrams} is similar to the well-known thermodynamic cycles in the context of combustion engines \cite{PhysRevA.15.2086}, where the enclosed area also represents the transacted energy per cycle.  By demonstrating the usefulness of \textit{eye-diagrams} in improving the design of our device, we seek to establish this  as a generic tool for wider application in the VEH community and beyond. 

\section{ Frequency response: Hidden energy branch}

\begin{figure}
\includegraphics[width=0.9\columnwidth]{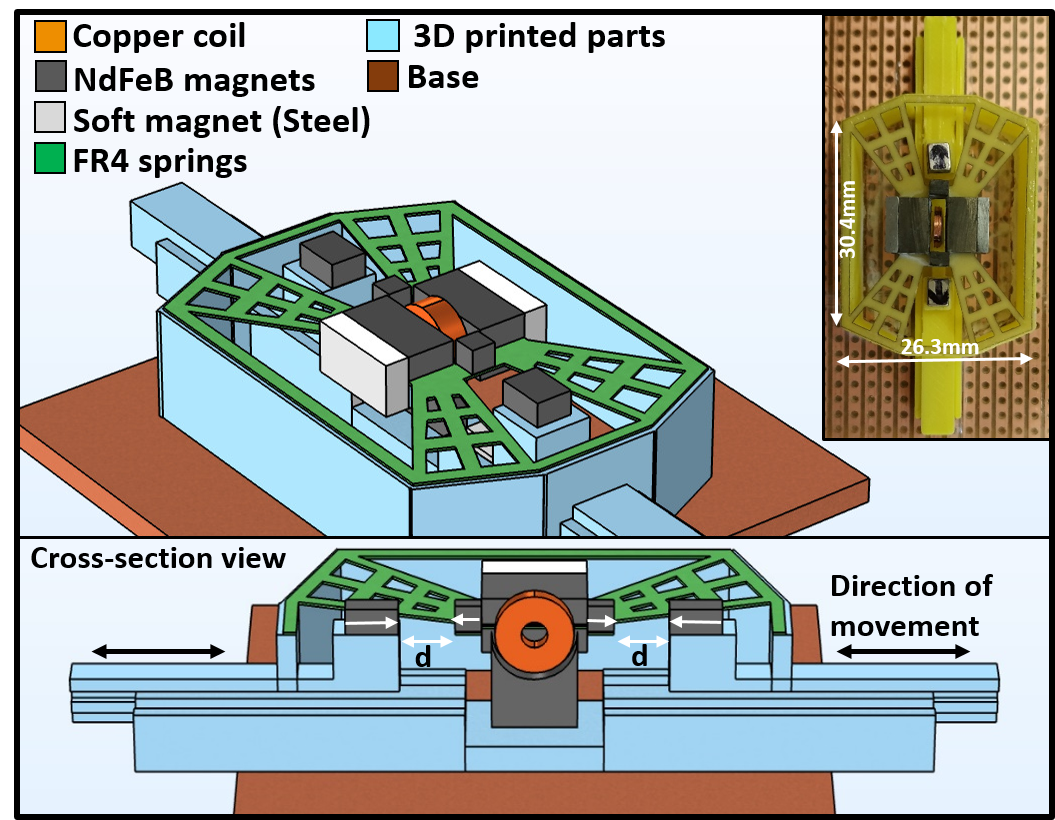}
\caption{\label{fig:schematic1} Schematic view (top) and  cross-section view (bottom) of the VEH. White arrows indicate the polarity of the magnets.The fabricated VEH prototype is shown at the inset.}
\end{figure}

\begin{figure*}
\includegraphics[width=1.6\columnwidth]{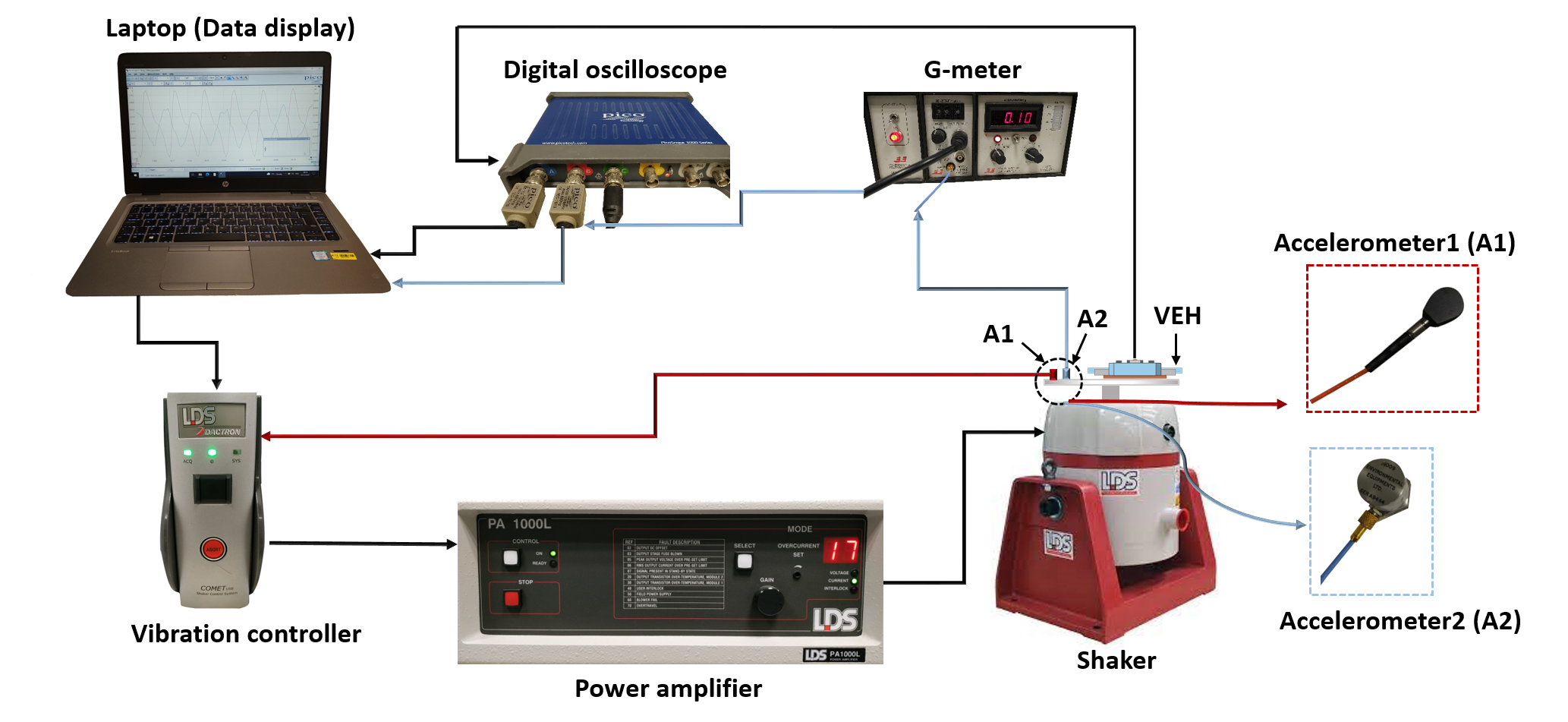}
\caption{\label{fig:schematic2} Schematic view of the experimental set-up for concurrent experimental measurement of the load voltage and the amplitude of excitation fed to the VEH to construct the \emph{eyes}.}
\end{figure*}

The employed tapered FR4 (Flame Retardant 4) spring architecture (laser micromachined), as shown in Fig.~\ref{fig:schematic1}, exploits the unique stress distribution \cite{paul2021tapered} arising from the tapered geometry to introduce a strong cubic nonlinear restoring force, while maintaining a small footprint of $1.23 cm^3$. Two pairs of repulsive permanent magnets, one pair fixed to the FR4 spring and the other mounted on movable rails are used  (bottom Fig.~\ref{fig:schematic1}) to  destabilize the central position of the load. Different parameters of the VEH are listed in Table~\ref{tab:table1}. The experimental set-up is shown in Fig.~\ref{fig:schematic2}. The electrodynamical characterization of the developed prototype has been performed with a Bruel and Kjær LDS V455 permanent magnet shaker that emulates real-world vibrations in a laboratory environment. The vibration of the shaker is controlled by an LDS Comet controller, and the output signal from the controller is fed to an LDS PA 1000L power amplifier. With a sweep rate of 1Hz/sec, the frequency of the excitation has been ramped up from 50Hz to 200Hz, and similarly swept back to 50Hz for different amplitudes of excitation (0.1g to 2.0g). A small piezoelectric CCLD accelerometer (DeltaTron 4517-002) placed near the harvester monitors the acceleration over frequency sweeps and feeds it back to the vibration controller (A1). Simultaneously, another accelerometer (A2) placed near the harvester monitors the amplitude of excitation and feeds to the g-meter (Environmental Equipments Ltd. Model 2025). The response from the harvester across an optimized load resistance $(2k\Omega)$ is recorded with a digital oscilloscope (Picoscope 3000 series). Concurrently, the output from the g-meter is recorded with the same oscilloscope. The 3D printed rails, as shown in Fig.~\ref{fig:schematic1} have been used to vary the distance $d$ between the repulsive sets of magnets. 

\begin{figure}
\includegraphics[width=1\columnwidth]{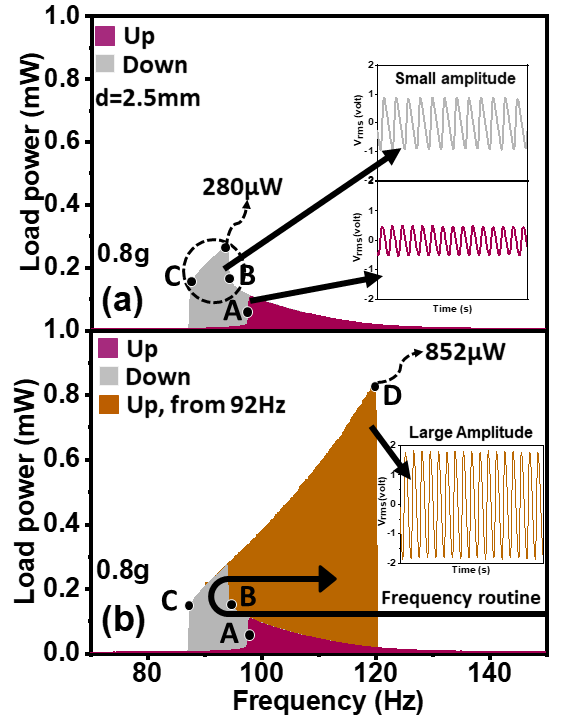}
\caption{\label{fig:schematic3} 
Variation of load power with (a) conventional up and down sweep and (b) designed frequency sweep of the drive for 0.8g excitation. The top and bottom inset of (a) are the time traces of the VEH's response on up and down sweep respectively. The inset of (b) shows the time trace of the VEH's response corresponding to large power output.}
\end{figure}

Fig.~\ref{fig:schematic3}(a) shows the variation of load power during driving frequency sweeps while the distance between the repulsive pairs of magnets and the amplitude of excitation are fixed at 2.5mm and 0.8g, respectively. As the driving frequency is swept up from 50Hz to 200Hz, the extracted power increases slowly (magenta) up to 0.1mW at point \textit{A}. Beyond this point, the power gradually decreases for higher driving frequencies. On sweeping the frequency of the drive down from 200Hz to 50Hz, the response jumps up at \textit{B}, and the VEH delivers more power across the load, maximizing up to 0.28mW (grey). The load power then reduces for lower driving frequencies and jumps down at the point \textit{C} to a low energy state. The top and bottom inset of Fig.~\ref{fig:schematic3}(a) shows the time trace of small amplitude oscillations exhibited by the VEH in the vicinity of \textit{A} and \textit{B}. 

\begin{table}
\caption{\label{tab:table1}
}
\begin{ruledtabular}
\begin{tabular}{lcdr}
\textrm{Title of parameters}&
\textrm{Values}\\
\colrule
Dimension of FR4 spring & $38 \times 23.3 \times 0.25 \text{mm}^3$\\
Dimension of coil slot & $8 \times 2 \text{mm}^2$ \\
Outer diameter of the coil & 6 mm \\
Inner diameter of the coil & 1 mm \\
Number of turns in coil & 2500 \\
Coil resistance\\ (measured) & 1032 $\Omega$ \\
Dimension of NdFeB N50 Magnets & $8 \times 4 \times 2 \text{mm}^3$\\
Dimension of Soft magnetic blocks & $8 \times 4.2 \times 1.6 \text{mm}^3$\\
Dimension of repulsive magnets \\fixed to FR4 & $2 \times 2 \times 2 \text{mm}^3$\\
Dimension of repulsive magnets \\on moving rail & $4 \times 2 \times 2 \text{mm}^3$\\
Electromagnetic coupling \\ (calculated) & 15 Wb/m \\
Inertial mass \\ (measured) & $3 \times 10^{-3} \text{kg}$ \\
Optimized load resistance \\ (measured) & 2 k$\Omega$ \\ 

\end{tabular}
\end{ruledtabular}
\end{table}

\begin{figure}
\includegraphics[width=1\columnwidth]{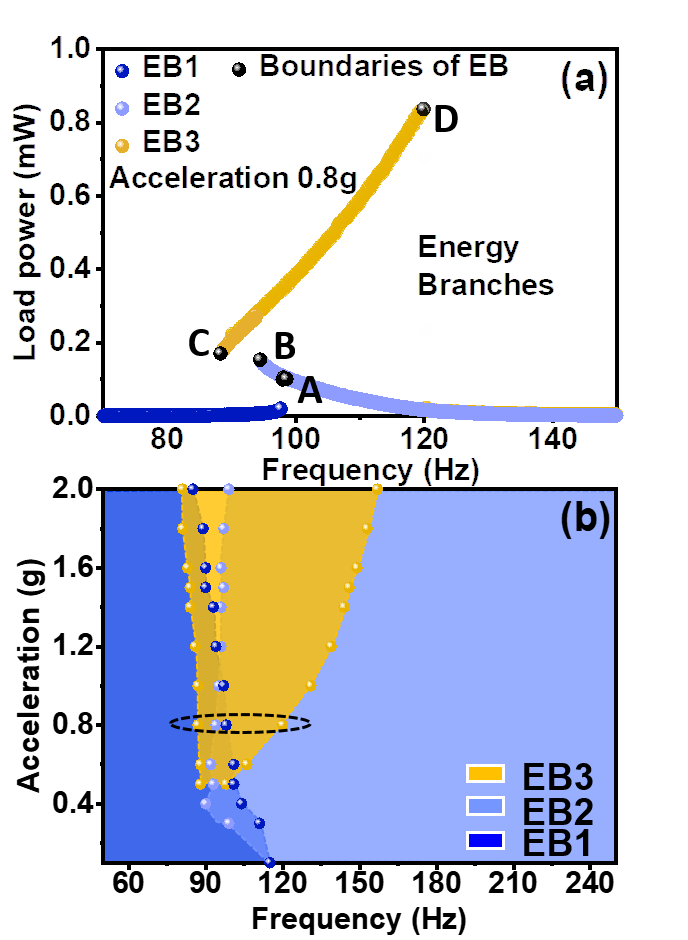}
\caption{\label{fig:schematic4} (a) Shows the extent of the energy branches EB1, EB2 and EB3 for 0.8g drive amplitude. The branch EB3 is achieved through the designed frequency routine. (b) Shows the mapping of energy branches EB1, EB2 and EB3 on the acceleration-frequency plane.}
\end{figure}

Interestingly, as shown in Fig.~\ref{fig:schematic3}(b), higher energy output can be achieved by designing a specific drive frequency schedule. While sweeping the frequency of the external drive down from 200Hz, instead of going all the way up to 50Hz, we turn the drive frequency up from 92Hz which is between the two jumping points \textit{B} and \textit{C}, as depicted by the arrow in Fig.~\ref{fig:schematic3}(b). During this up sweep, the VEH now delivers large power output (brown) of up to 0.85mW at the point \textit{D} before falling down to a low energy state. This high output is not achieved in the simple up and down sweep as shown in Fig.~\ref{fig:schematic3}(a) and it is a consequence of the existence of multiple stable energy branches that have been shown later in Fig.~\ref{fig:schematic4}(a). More explicitly, there is a low energy branch EB1 (dark blue) which is the only stable branch at low frequencies and it terminates at the point \textit{A}. Then there is an intermediate energy branch EB2 (light blue), the only stable branch at the high frequencies which terminates at \textit{B}. Finally, there is a high energy output branch EB3 (yellow) which extends from \textit{C} to \textit{D}, and can only be selected through particular frequency schedule of the drive. It is interesting to note that, between the points \textit{B} and \textit{A}, all the three energy branches co-exist, and the frequency schedule  of the drive determines which of the three branches is selected. 

The energy branches depends also on the acceleration of the drive. In Fig.~\ref{fig:schematic4}(b), we show the extent of the energy branches in the acceleration-frequency plane (for d =2.5mm). The dot symbols mark the experimentally obtained boundaries of the respective energy branches. The linear frequency of the oscillator is at 115Hz for very low acceleration (0.1g) of the external drive. With increasing acceleration, the two energy branches EB1 and EB2 overlap and form a hysteresis region of up to 14Hz at 0.4g. Then at a drive of 0.5g the high energy branch EB3 starts to emerge (yellow region in Fig.~\ref{fig:schematic4}(b)). As a consequence of its position in the overlap region between EB1 and EB2, the branch EB3 can only be achieved by following the frequency schedule as explained before. This energy branch (EB3) extends up to 76Hz as the external drive increases to 2g and provides the largest energy output of all branches. It should be noted that the high energy branch EB3 aids the system to generate more energy over a considerably wider bandwidth of operable frequencies which makes it a potential candidate for harnessing mechanical energy from broadband vibrations. However, the multistable characteristics of this VEH makes it difficult to achieve and sustain the high energy state consistently. Using controlled electrical actuation \cite{mallick2016surfing} is a viable route to switch the state of this system to higher energy state, while enabling the VEH to capture substantial mechanical energy from real-world vibrations.  

\section{Reduced Order Model for the VEH}

\begin{figure}
\includegraphics[width=1\columnwidth]{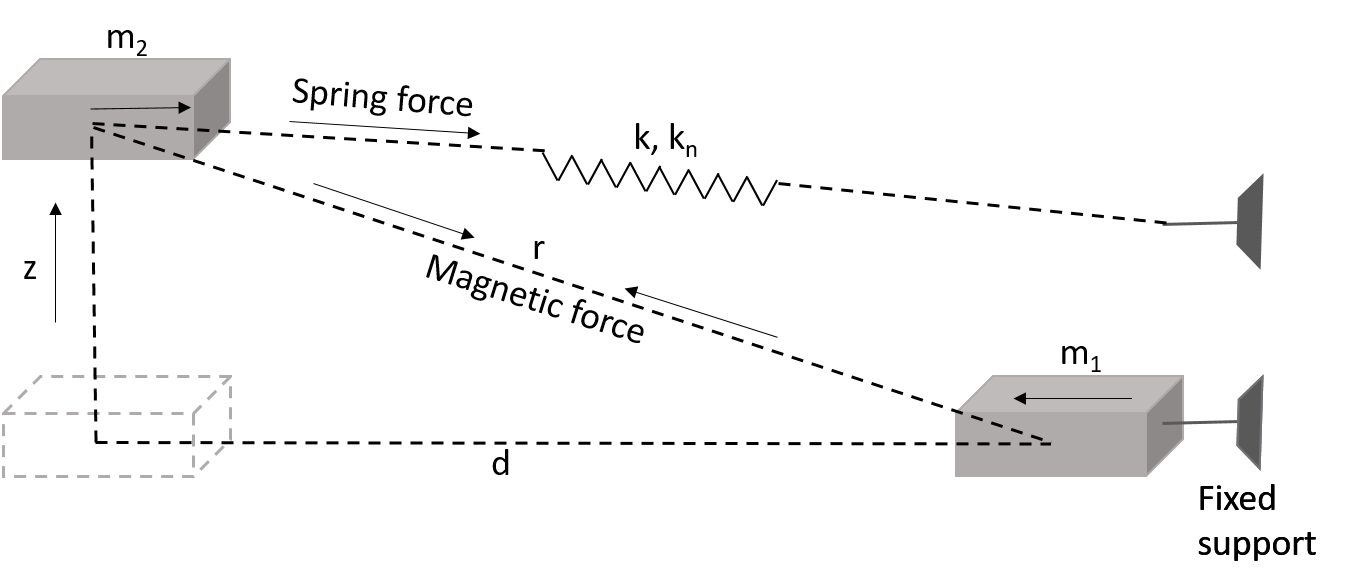}
\caption{\label{fig:schematic5} Schematic of the magnetic interaction; for the sake of simplicity only the vertical displacement of $m_{2}$ is considered here.}
\end{figure}

To describe the dynamics of the nonlinear VEH system, let us consider the following general equation of motion for our driven system,
\begin{eqnarray}\label{eq_motion}
m\ddot z + c\dot z + \gamma I + \frac{\partial {U (z)}}{\partial z} = F \sin{\omega_0\; t}
\end{eqnarray}
here, $z$ is the vertical displacement of the moving magnet and $m$ is its mass; $c$ is the mechanical damping parameter, $\gamma$ is the electromagnetic coupling factor, $U(z)$ is the potential energy, and $F \sin{\omega_0 t}$ is the external drive.The interaction between the coil and the magnet has been emulated using the finite element analysis tool Ansys Maxwell \cite{paul2021tapered}. The electromagnetic coupling factor $\gamma$ represents the spatial gradient of magnetic flux that is experienced by the coil under consideration, which has been calculated to be 15mWb/m in this case.
$I$ is the current through the load resistor which is expressed as,
\begin{eqnarray}
I=\frac{\gamma \dot z}{R_C+R_L}
\end{eqnarray}
where, $R_L$ is the load resistor and $R_C$ is the coil resistor.  To model the effect of the external magnet, we consider the repulsive interaction between two magnets with magnetic dipoles $m_2$ on $m_1$ at a distance $d$, as shown in Fig.\ref{fig:schematic5}. The potential energy $U(z)$ for this interaction is given by  \cite{griffiths2005introduction},
\begin{eqnarray}
U(z)&=& \frac{\mu_0}{4 \pi (z^2+d^2)^{3/2}}{\left[\frac{3 m_1 m_2 d^{2}}{(z^2+d^2)}- m_1 m_2 \right]}\\
&=&\frac{\mu_0 m_1 m_2 }{4 \pi d^3} {\left[\frac{2 - \frac{z^2}{d^2}}{(1+\frac{z^2}{d^2})^{5/2}} \right]}\nonumber
\end{eqnarray}

Taking into account the contribution from the repulsive set of magnets as well as the linear and nonlinear spring force arising from the spring bending and stretching respectively ($k$ and $k_n$ being the linear and nonlinear spring stiffness coefficient), the total force ${F_ {overall}(z)}$ arising from the spring and the magnets can be expressed as,
\begin{eqnarray}
F_{overall}= k z + k_n z^3- \frac{\partial U (z)}{\partial z}\\ \nonumber
= k z + k_n z^3+F_{mag}(z)\nonumber
\end{eqnarray}
Then the equation of motion of the VEH under consideration takes the form
\begin{eqnarray}
\ddot z + \frac{c}{m}\dot z +\frac{\gamma^2}{m (R_C + R_L)} \dot z\\+ \frac{k}{m} z + \frac{k_n}{m} z^3+ \frac{F_{mag}(z)}{m} 
&=& \frac{F}{m} \sin{\omega_0\; t}\nonumber
\end{eqnarray}
It is convenient to non-dimensionalise this equation. First we choose the dimensionless time parameter $\tau = \omega t$, 
\begin{eqnarray}
\frac{d}{d t}&=& \frac{d\tau}{d t}\frac{d}{d\tau}= \omega\frac{d}{d\tau}
\end{eqnarray}
Substituting this in the equation of motion and selecting $\omega=\sqrt{\frac{k}{m}}$, the equation of motion takes the following form, 
\begin{eqnarray}
\frac{d^2 z}{d\tau ^2}  + \frac{c}{m} \sqrt{\frac{m}{k}} \frac{d z}{d\tau} +\frac{\gamma^2}{m (R_C + R_L)} \sqrt{\frac{m}{k}}  \frac{d z}{d\tau}\\ \nonumber + z + \frac{k_n}{k} z^3+ \frac{F_{mag}(z)}{k}  = \frac{F}{k} \sin{\hat{\omega}\tau}\\ \nonumber
\end{eqnarray}
Scaling $z$ such that $\hat{z} = z a$, choosing $\frac{k_n}{k}=a^2$, and combining the two damping terms using $\frac{c}{\sqrt{mk}} +\frac{\gamma^2}{\sqrt{mk} (R_C + R_L)} =D_{total}$, the simplified form of the equation of motion is,
\begin{eqnarray}
 \frac{d^2\hat{z}}{d\tau ^2}  + D_{total}\frac{d \hat{z}}{d\tau} + \hat{z}
+ \hat{z^3} \nonumber\\
 +\frac{\mu_0 m_1 m_2 a }{4 \pi m d^3 k}  \left[\frac{-2 \frac{\hat{z}}{a}}{d^2 (\frac{\hat{z}^2}{a^2 d^2} +1)^{5/2}}-\frac{5 \frac{\hat{z}}{a} (2 - \frac{\hat{z}^2}{a^2 d^2})}{d^2 (\frac{\hat{z}^2}{a^2 d^2} +1)^{7/2}} \right] \label{ROM1}\\
= \frac{F a}{k}\nonumber \sin{\hat{\omega}\tau}
\end{eqnarray}
We further introduce the nondimensionalized parameters as $\hat{d}= a d$, $\hat{F}=\frac{Fa}{k}$ and $(\frac{3}{4 \pi}) ( \frac{\mu_0 m_1 m_2}{k}) =\frac{\hat{p}}{a^5}$. We then get the following dynamical equation for the Reduced Order Model (ROM),
\begin{eqnarray}
\frac{d^2\hat{z}}{d\tau ^2}  + D_{total}\frac{d \hat{z}}{d\tau} + \hat{z}
+ \hat{z^3}  \label{ROM2} \\
 +\hat{p} \frac{\hat{z}}{\hat{d^5}}\left[\frac{{(\frac{\hat{z^2}}{\hat{d^2}}-4) }}{(\frac{\hat{z}^2}{\hat{d}^2}+1)^{7/2}}\right]
&=& \hat{F}\nonumber \sin{\hat{\omega}\tau}
\end{eqnarray}

Equation \eqref{ROM2} is solved by using the fourth-order Runge-Kutta method in MATLAB. We will now utilize this nondimensionalized model to investigate the complex dynamics associated with the presented system. The linear and nonlinear parameters that have been used in this model are shown in Table~\ref{tab:table2}. The linear and nonlinear spring stiffness coefficients of the spring structure have been estimated using the solid mechanics module of COMSOL Multiphysics platform. The stationary analysis in COMSOL is used to excite the spring structure at the fundamental vibrational frequency of 94Hz while sweeping the external force from -2N to 2N and recording the displacement of the spring. The spring stiffness coefficients are extracted from the force-displacement relationship. The mass, coil and load resistance are measured and are fed into the reduced order parameters.

\begin{table*}
\centering
\caption{\label{tab:table2}%
Parameters and their values}
 \begin{tabular}{| c | c | c | c |} 
 \hline
Parameter & Expression & Unit & Value  \\

 \hline\hline
 
 k & spring stiffness & N/m & 2085.6 \\
 
 $k_n$ & nonlinear spring stiffness & $N/m^3$ & 21.5 $\times 10^9$ \\
 
 m & mass & kg & 3$\times 10^{-3}$ \\
 
 a & $\sqrt{\frac{k_n}{k}}$ & $m^{-1}$ & $3.2152\times 10^3$ \\
 
 $\omega_0$ & $2 \pi f_0$ & $s^{-1}$ & 590.61\\
 
 c & $2 m \omega_0 \rho_m $ & $Ns/m$ & 0.0106\\
 
 $R_C$ & Coil Resistance & $\Omega$ & 1032\\
 
 $R_L$ & Load Resistance & $\Omega$ & 2000\\
 
 $\gamma$ & Electromagnetic coupling & Wb/m & 15\\
 
 $D_{total}$ & $\frac{c}{\sqrt{mk}} + \frac{\gamma^2}{\sqrt{mk} (R_C + R_L)}$ & Dimensionless & 0.0042+0.0297= 0.034 \\

$d$ & Distance between magnets & mm & 2.5 (changes between 1mm to 7mm)  \\

$\hat{d}$ & $d \times a$ & Dimensionless &  $\sim 8$\\

$\mu_0$ & Free space permeability & H/m & 12.57 $\times 10^{-7}$\\

$m_1$ & Magnetic moment & $Am^2$ & 12 $\times 10^{-3}$\\

$m_2$ & Magnetic moment & $Am^2$ & 7.5 $\times 10^{-3}$\\
  
$\hat{p}$ & $(\frac{3}{4 \pi}) ( \frac{\mu_0 m_1 m_2}{k})\times a^5$ & Dimensionless & $4.16\times 10^{3}$ \\
 
 $\hat{F}$ & $\frac{Fa}{k}$ & Dimensionless & 0.037  \\
   &  &  & (F=$8m/s^2$) \\
 
 \hline
 \end{tabular}
 
\end{table*}

We use the ROM to investigate the dynamic response of the system. The magnitude of the applied force and the distance between the magnets are kept fixed at 0.86g and 2.0765mm respectively. As shown in Fig.6, with traditional up and down sweep of frequency, a sudden jump is observed at $\hat{\omega}=0.6$, which is highlighted as point $B$. On designing a frequency routine as explained before to achieve the higher energy state, we observed a similar high energy response from the system as the drive frequency is swept up from 0.56 (point $C$) to 2.2 (point $D$). Similar to the previously explained experimental observation, we can also notice here a low energy branch EB1 that terminates at $A$, the higher energy branch EB2 that terminates at $B$, and the hidden energy branch EB3, which is achieved through the designed frequency routine, extends from $C$ to $D$. Hence, we conclude that the ROM is able to reproduce the previously presented experimental results.

\begin{figure*}
\includegraphics[width=1.4\columnwidth]{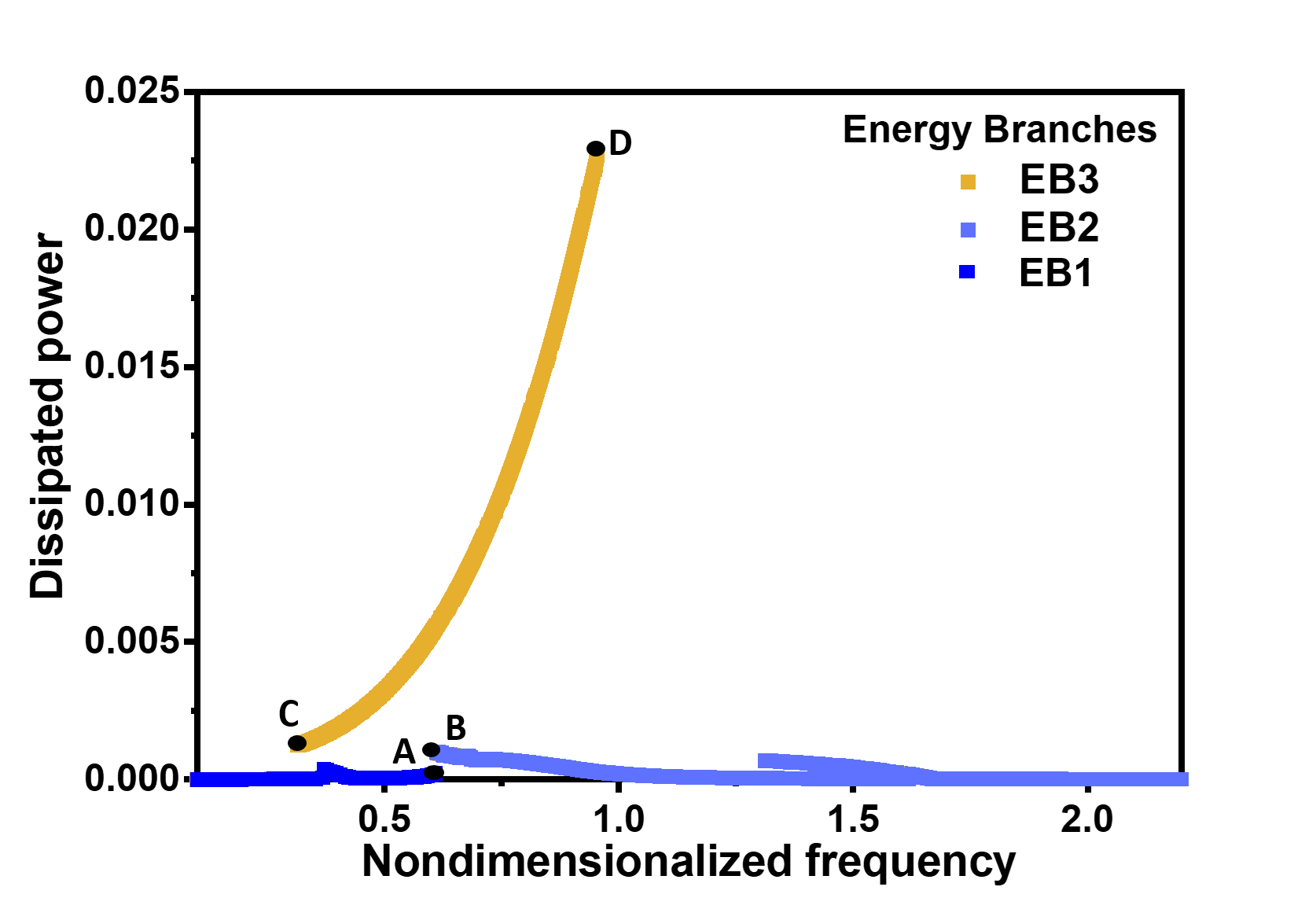}
\caption{\label{fig:schematic6} Plot of the dynamic response of the system using the ROM showing the extent of the different energy branches.}
\end{figure*}

\section{Eye diagrams}

In order to investigate the dynamical features of the branches, and in particular to compare their energy output, we now introduce the concept of \emph{eye diagrams}, which provide intuitive insight into the various dynamical states on the basis of  experimentally observable quantities. 

Revisiting the equation of motion \eqref{eq_motion}, let us now assume that the period of the solution $z(t)$ equals the period of the external drive $T= 2\pi/\omega_0$ and let us further assume that at a time $T_0$ the displacement $z(T_0)$ is at a maximum. Multiplying both sides of equation \eqref{eq_motion} by $\dot z$ and integrating over one period $T$ leads to the condition,
\begin{align}
E_m+E_e = E_f,\label{Ed_int}
\end{align} with
\begin{align}
E_m &= \int\limits_{T_0}^{T_0+T} c\; \dot{z}^2\; dt, \quad
E_e = \int\limits_{T_0}^{T_0+T} \gamma I(t)\; \dot{z}\; dt, \\
E_f &= \int\limits_{T_0}^{T_0+T} F \sin{(\omega_0\; t)\; \dot z\; dt} .
\end{align}
\begin{figure}
\includegraphics[width=0.8\columnwidth]{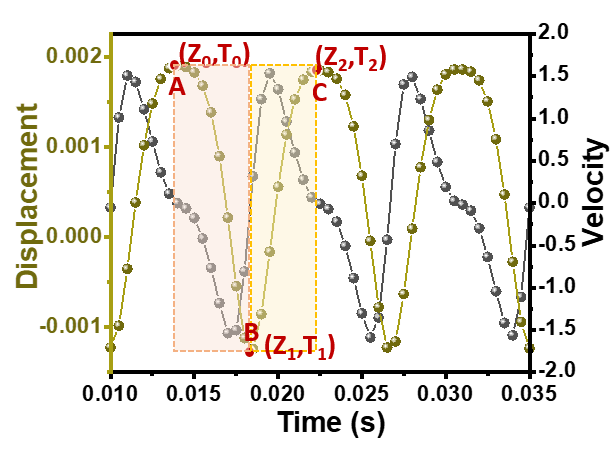}
\caption{\label{fig:schematic7} 
Plot of a nonlinear displacement and velocity as a function of time. The point \text{A} and \text{B} represents the local maxima, and \text{C} represents the local minima of the displacement function.}
\end{figure}
\begin{figure}
\includegraphics[width=0.85\columnwidth]{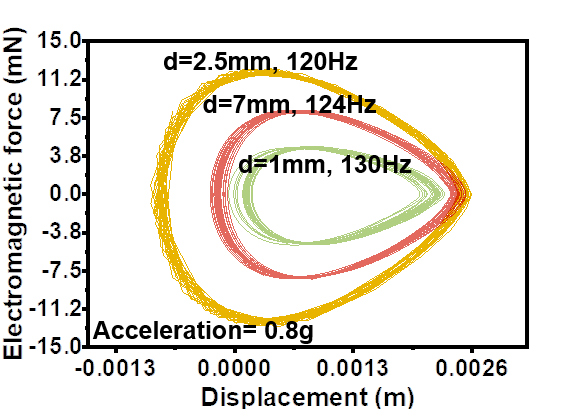}
\caption{\label{fig:schematic8} 
The \textit{eyes} corresponding to the electromagnetically transduced energy for discrete values of d(=1mm, 2.5mm, 7mm) are shown.}
\end{figure}
Here, $E_m$ is the mechanical contribution to the dissipated energy over one period and $E_e$ is the electromagnetically transduced energy over one period of the drive. $E_f$ is the energy injected through the external drive.
\begin{figure*}
\includegraphics[width=2.1\columnwidth]{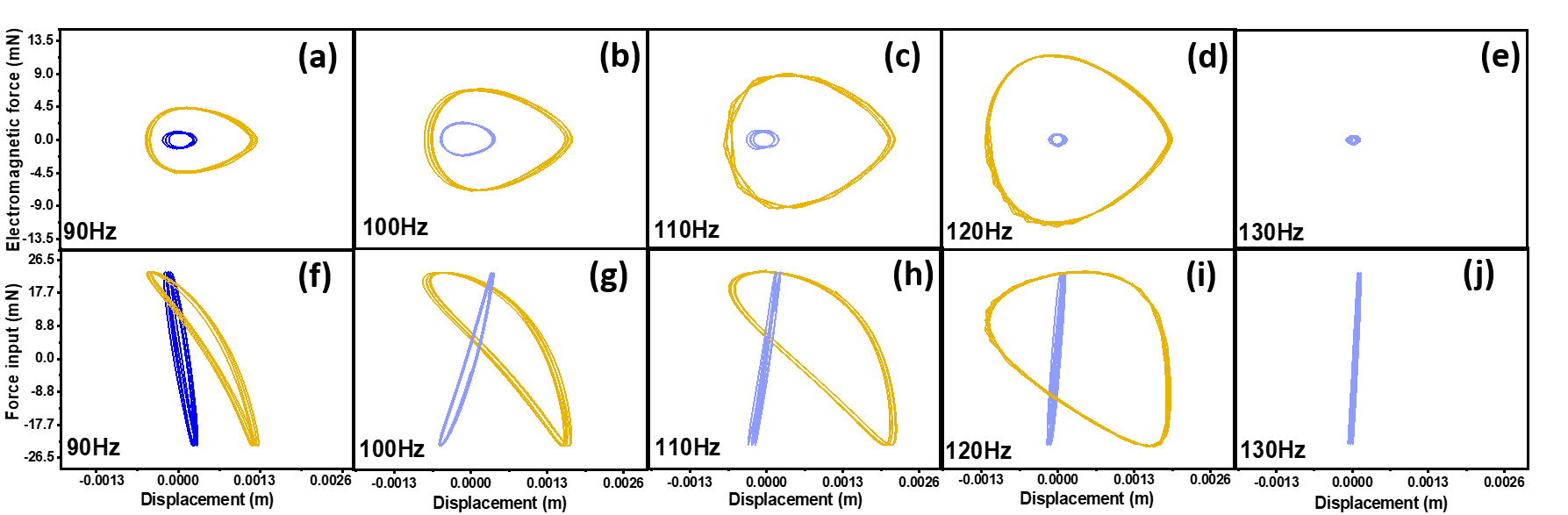}
\caption{\label{fig:schematic9}With d=2.5mm, the \textit{eyes} corresponding to electrical energy dissipating through involved damping are shown in (a)-(e). The \textit{eyes} representing mechanical energy injection into the VEH are shown in (f)-(j). The colours of these \textit{eyes} correspond to the different energy branches depicted in Fig.2(a) and Fig.3.}
\end{figure*}
\begin{figure}
\includegraphics[width=1\columnwidth]{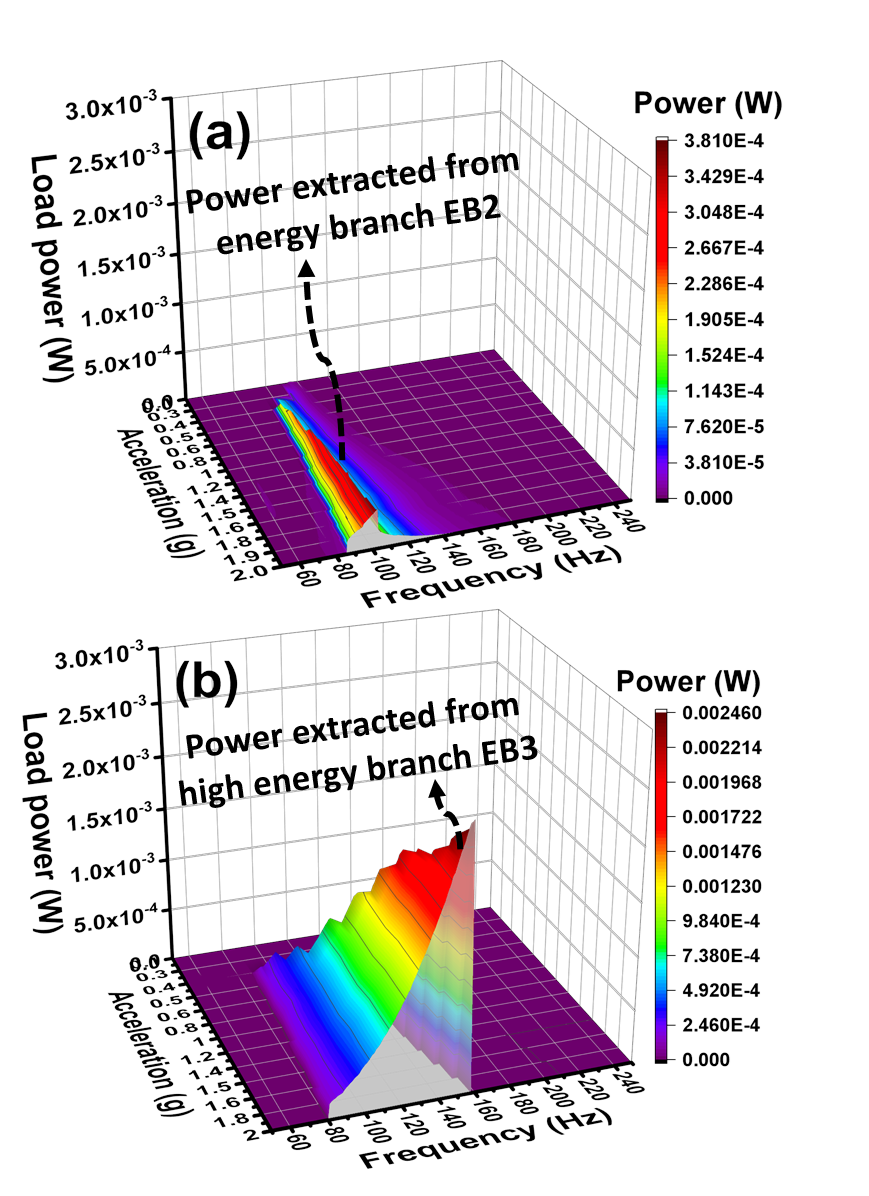}
\caption{\label{fig:schematic10}  Variation of the deliverable power of the VEH as a function of the frequency and amplitude of the drive for d =2.5mm, (a) with conventional up and down frequency sweep and (b) with specifically designed frequency routine.}
\end{figure}
Equation \eqref{Ed_int} represents the energy balance of these energies associated with the energy harvesting system.  Motivated by similar representations of the energy balance in combustion engines, let us now represent the two sides of equation \eqref{Ed_int} as an enclosed area in a suitable force-displacement plane. We do this by substituting the integration in time $t$ by an integration in the displacement $z$.  Since $z(t)$ is a non-monotonous function, the substitution is split up into time intervals $[T_{k-1},T_k]$ where, $T_k$ is a maximum (minimum) of $z(T_k)$ for even (odd) $k$ for $k=0,\ldots, 2n$, where $n$ is the number of minima per period $T$.  This also implies $T_{2n}=T_0+T$. This is illustrated for the case $n=1$ in Fig.~\ref{fig:schematic7}). Defining $\hat{t}_k(z)$ as the inverse function of $z(t)$ on the interval $[T_{k-1},T_k]$ then yields
\begin{eqnarray}
  E_{x} &=& \sum_{k=1}^{2n} \int_{Z_{k-1}}^{Z_{k}}  F_{xk}(z)dz,  \label{Exarea}
\end{eqnarray}
where the subscript $x$ refers to either $m$, $e$ or $f$ and $F_{mk}(z)=c\dot{z}(\hat{t}_k (z))$, $F_{ek}(z)=\gamma I(\hat{t}_k (z))$, $F_{fk}(z)=F\sin(\omega_0 \hat{t}_k (z))$ are the corresponding forces.  Equation \eqref{Exarea} shows that the energies $E_m$, $E_e$ and $E_f$ appearing in \eqref{Ed_int} can be interpreted intuitively as the areas enclosed by the functions  $F_{mk}(z)$ , $F_{ek}(z)$ and $F_{fk}(z)$ in a displacement versus force diagram.   

\section{Experimental Methods, Results and Discussions}

To experimentally determine the function $F_{ek}(z)$, the electromagnetic force, we measure the VEH's output voltage $V_L$. This allows us to obtain $\dot{z}= V_L  (R_c+R_L)/\gamma R_L$. We then calculate the displacement $z$ by integrating the velocity $\dot{z}$. 
Fig.~\ref{fig:schematic8} depicts the resulting force displacement diagram for fixed acceleration and various values of the interspacing between the repulsive magnets (\text{d}). The enclosed area corresponds to the energy $E_e$ transduced in one period. As this shape resembles the shape of an \textit{eye}, we call this an \emph{eye diagram}. The eye for the VEH topology with $d=2.5mm$ encloses the largest area among all eyes and therefore represents the largest energy transaction into the electrical domain per forcing period.

\begin{table}
\caption{\label{tab:table3}%
Energy conversion ratios from mechanical to electrical domain at different energy branches
}
\begin{ruledtabular}
\begin{tabular}{lcdr}
\textrm{Frequency}&
\textrm{EB1}&
\textrm{EB2}&
\textrm{EB3}\\
\colrule
90Hz & 0.13 & - & 0.91 \\
100Hz & - & 0.52 & 0.64 \\
110Hz & - & 0.43 & 0.62 \\
120Hz & - & 0.1 & 0.52 \\
\end{tabular}
\end{ruledtabular}
\end{table}
Further, to experimentally determine the function $F_{fk}(z)$, we measure the external acceleration fed to the oscillator using a piezoelectric accelerometer attached to the base of the excitation source and we multiply this with the mass of the system ($3 \times 10^{-3}\text{kg}$) to obtain $F_{fk}(z)$. The displacement is obtained from the electromotive force measurement as explained before. An example is shown in Fig.~\ref{fig:schematic9}(f) for the two branches EB1 and EB3. The area enclosed in the $F_{fk}-z$ plane stands for the amount of mechanical energy fed to the VEH from the drive. Due to the energy balance in equation \eqref{Ed_int}, this corresponds to the area in the $F_{ek}-z$ plane as shown in Fig.~\ref{fig:schematic9}(a). In Fig.~\ref{fig:schematic9}(a)-(j) we compare the eye diagrams in the $F_{ek}-z$ plane with the corresponding diagrams in the $F_{fk}-z$ plane for various frequencies. We observe that the shape of the eye evolves differently for the three different branches that are shown Fig.~\ref{fig:schematic4}(a). In particular the branch EB3 corresponds to large areas enclosed in  Fig.~\ref{fig:schematic9}(d) and (i), while the co-exisiting branch EB2 only encloses a small area. 
The area enclosed by the eye corresponding to this high energy branch EB3 in Fig.~\ref{fig:schematic9}(i) represents $98\mu J$ mechanical energy that is fed into the nonlinear VEH through the external excitation ($E_f$). On the other hand, the area enclosed by the eye for EB3 in Fig.~\ref{fig:schematic9}(d) depicts the fraction of this mechanical energy, $51\mu J$, that is transacted into electrical domain by the VEH ($E_e$) per cycle. Now we define the energy conversion ratio from mechanical to electrical domain as, 
\begin{eqnarray}
Energy \; Conversion \; Ratio =\frac{E_e}{E_f}
\end{eqnarray}

It is important to note that this energy $E_e$ is dissipated in both the coil and the load resistance; only the fraction that is dissipated across the load resistor represents the usable energy which could be utilized in a target application.The energy values as mentioned above, obtained from the area of the \emph{eyes} reflects an energy conversion ratio of 0.52. On the other hand, the eye corresponding to EB2 in Fig.\ref{fig:schematic9}(i) represents only $2.5\mu J$ mechanical energy acquired from the external force, only $0.3\mu J$ of this energy gets transacted as usable electrical energy, yielding a conversion efficiency of 0.12. Similarly, the EB1 only converts $0.9\mu J$ energy into the electrical domain, a fraction of $8\mu J$ energy that the drive provides to the VEH, resulting a conversion efficiecy of only 0.11. Interestingly, this efficiency increases to 0.91 when the energy contribution from the energy branch EB3 is taken into account. The energy conversion ratio corresponding to each energy branches for the 90Hz, 100Hz, 110Hz and 120Hz drive frequency has been summarized in Table~\ref{tab:table3}. 

Furthermore, the shape of the eye diagrams for EB3 deviate strongly from the simple ellipse, which is characteristic for a harmonic oscillator.  We therefore conclude that EB3, which is only obtained through a special frequency schedule, is inherently connected to the nonlinear force in our system. The eye diagrams are therefore a useful tool to experimentally explore the nonlinear character of the various co-existing branches. 

To connect to the well studied linear case, let us consider the shape of the \textit{eyes} corresponding to the energy branch EB1 in Fig.~\ref{fig:schematic9}(a) and (f). They are close to the shape of an ellipse, which suggests that the VEH performs harmonic oscillations, similar to a simple linear harmonic oscillator.  In this context, the question arises, if we could not simply define an appropriate \emph{phase} which is able to characterize the response of the system.  In fact, the phase is a useful tool for linear oscillators where the periodic response to an external drive of the form $F(t) = \sin(\omega t)$ only consist of a single harmonic component, i.e. $z(t) = z_0 \sin(\omega t - \phi_0)$. In this case the quantity $\phi_0$  uniquely defines the \emph{phase} of the response, which can also be used to characterise the energy transaction in the linear case.  However, in a \emph{nonlinear} system multiple frequency components are present in the response, i.e $z(t) = z_0 \sin(\omega t- \phi_0) + z_1 \sin(2 \omega t - \phi_1) + \ldots$, and the relationship between drive and response cannot be expressed in terms of a single phase.  In this case, the eye-diagrams introduced before prove more useful, as they take into account all frequency components of the response. In Appendix-A we provide a simple example which shows that in a nonlinear oscillator the energy transaction depends on the phases of higher frequency components. \\

In this case, the phase difference between displacement and the input excitation determines the enclosed area and thereby the energy transacted during one cycle. This corresponds to the well known role of the phase in the linear oscillator, where a phase difference of $\pi/2$ corresponds to the peak of the resonance. On the other hand, all of the \textit{eyes} from the high energy state EB3 have asymmetrical shape, which indicates that the strong nonlinear restoring force arising from the the stretching of the spring as well as from the repulsive magnetic interactions predominates here. Similarly, the energy state EB2 possess very little asymmetry in the \textit{eyes} corresponding again to weak nonlinearities. 

As discussed above, the shape of the eyes are different for each energy branch.  This fact can be exploited for the discovery of previously unknown branches.  Let us for example revisit the frequency down-sweep in Fig.~\ref{fig:schematic3}(a).  In this case, the small jump in the load power at point B reveals the presence of another branch. Such a jump is however not guaranteed to be visible in all cases where a transition between branches occurs. In contrast, if we consider the transition between the light blue eye in Fig.~\ref{fig:schematic9}(g) for 100Hz  and the yellow eye in Fig.~\ref{fig:schematic9}(f) for 90Hz, we see that the shape and orientation of the eye markedly changes in addition to the enclosed area.  This provides therefore a much stronger signal for a branch change and in this case prompts us to further explore the hidden branch EB3, which turns out to feature the largest energy output available in this device. 

In order to illustrate the difference in power output obtainable for given input frequency and acceleration, let us consider Fig.~\ref{fig:schematic10}(a) and (b). In
Fig.~\ref{fig:schematic10}(a) we show the load power as the drive frequency is simply swept down from 250Hz to 50Hz, while the acceleration is fixed for each sweep.  On the other hand, in Fig.~\ref{fig:schematic10}(b) the power following the specific frequency schedule as explained before to achieve higher energy states is shown. We observe that in the parameter regime, where the branches EB2 and EB3 overlap, the branch EB3 has a much higher energy output which is reached in Fig.~\ref{fig:schematic10}(b) but not Fig.~\ref{fig:schematic10}(a). 
For example, at 0.5g drive amplitude, the peak load power approximately doubles from 0.18mW to 0.33mW, when the system follows EB3 instead of EB2. The delivered power increases to 1.3mW for 1g, with a 44Hz bandwidth. As shown in Fig.~\ref{fig:schematic10}(b), the load power further increases with increasing acceleration up to 2.8mW while offering a large bandwidth of up to 76Hz at 2g. This wide operable frequency bandwidth corresponding to the branch EB3 is a feature due to the nonlinearity of our device. This offers a unique benefit for harnessing real-world vibrations where no apriori knowledge of the prevalent frequency components is available. In the weakly excited regime, this VEH essentially behaves as a linear resonator. The obtainable frequency bandwidth is as low as 3.19Hz for 0.1g excitation which restrains the efficiency for harvesting energy from brodaband vibrations. Since the bandwidth is a key performance metric, this outlines the advantage of employing a nonlinear device compared to a linear one, which despite of meeting the resonance condition offers a low frequency bandwidth. 

\begin{figure}
\includegraphics[width=0.85\columnwidth]{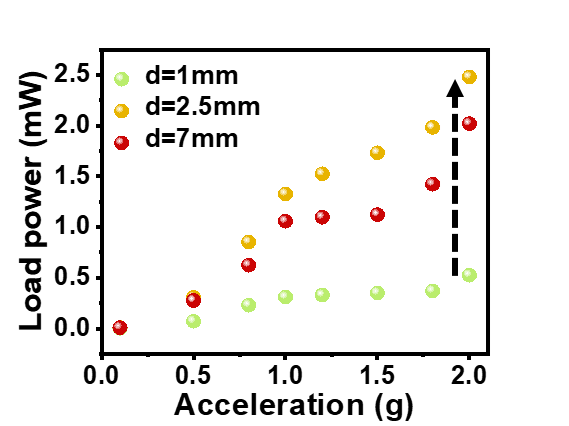}
\caption{\label{fig:schematic11}The performance comparison of the VEH topologies in terms of the extracted power across a resistive load ($2k\Omega$) for different values of interspacing between the repulsive magnets \emph{d}(=1mm, 2.5mm, 7mm).}
\end{figure}
Fig.~\ref{fig:schematic11} shows a comparison of peak load power obtained from this VEH for different values of the interspacing between the repulsive magnets ($d$). When the magnets are very close ($d$=1mm), the strong repulsive force between results in very low electrical power generation. For example, for a 1g drive, the extracted peak load power is only 0.32mW. In contrast, if the  magnets are at a large distance (d=7mm), the VEH exhibits larger oscillations about the equilibrium state. This improves the deliverable power to 1mW. As the magnets are placed at an intermediate value of d=2.5mm, the associated nonlinear effects both from the spring stretching and the magnetic interaction result in a large peak load power of 1.3mW that is extracted from the energy branch EB3. This corresponds to approximately 30\% and 300\% improvement in the power outcome as compared with the VEH topology with d=7mm and d=1mm, respectively. Therefore, this demonstrates the effectiveness of using a repulsive pair of magnets at an optimized distance to enhance the overall performance of the VEH.

\section{Conclusion}

To summarize, a wideband vibration energy harvester is presented with multiple nonlinear force acting on the system that gives rise to a number of energy branches. Some of these branches are ``hidden'' in the sense that they are not fully reached by simple frequency up or down sweeps. We designed a particular frequency schedule to reach those branches which substantially improved the energy output. The different branches have been experimentally characterised through \emph{eye diagrams}, which directly illustrate the magnitude of the transacted energy per cycle. The energy harvesting device yields 1.3mW power (at 1g) across a suitable load resistor providing an enhanced operable frequency bandwidth of 44Hz. This energy harvesting system transduces mechanical energy into usable electrical energy at a conversion efficiency of 52\%.

\begin{acknowledgments}

The author would like to thank Tony Compagno for the help in executing the experiments and the useful discussions. This work is financially supported by a research grant from Science Foundation Ireland (SFI) and is co-funded under the European Regional Development Fund Grant Number 13/RC/2077. This is also part funded by the EU-H-2020 project ‘Enables’, Project ID: 73095 and the Science Foundation Ireland (SFI) Frontiers for the Future Programme (FFP) Award Grant (Grant ID: 21/FFPA/10003).
\end{acknowledgments}

\appendix
\section{}
\label{appendix_I}
To explicitly highlight the contrast of the role of phase in linear and nonlinear systems, we here provide a mathematical example. Let us first consider a linear system that is driven with harmonic excitation of the form  $F(t) = F_0 \sin(\omega t)$ and the response of the system is expressed through the displacement $z(t) = z_0 \sin(\omega t - \phi_0)$. where $\omega$ is the frequency and $\phi_0$ is the phase difference between the applied force and the response of the system. We can find the energy that is injected into the system through the external drive,
\begin{align}
E_{linear} &= \int\limits_{0}^{T} F(t)\; \dot{z}(t)\; dt\\
&= \frac{F_0 \; z_0 \; \omega}{2} \int\limits_{0}^{T} \left[\sin(2 \omega t - \phi)\; + \sin (\phi)\;\right] dt \nonumber\\
&= \frac{F_0 \; z_0 \; \omega T}{2} \sin(\phi)\nonumber
\end{align}
This shows the contribution of phase $\phi$ in controlling the energy transaction and hence the performance of such a linear system. 

Let us now consider a nonlinear system with displacement of the form $z(t) = z_0 \sin(\omega t - \phi_0) +z_1 \sin(2\omega t - \phi_1)$ which comprises higher harmonic components along with different phases $\phi_0$ and $\phi_1$. Now, considering the same external forcing as that of the linear system, the energy fed into this nonlinear system can be expressed as,
\begin{align}
E_{nonlinear} 
&= \int\limits_{0}^{T} F_0 \;\sin(\omega t)\; \left [\omega z_0  \cos(\omega t - \phi_0) \right] \\
& + \int\limits_{0}^{T} F_0 \; \left [ 2\omega z_0 \cos(2\omega t - \phi_1)\; \right ] dt \nonumber\\
&= \frac{F_0 \; z_0 \; \omega}{2} \int\limits_{0}^{T} \left[sin(2 \omega t - \phi_0)\; + sin (\phi_0)\;\right] dt \nonumber\\
+&  \frac{2 F_0 \; z_0 \; \omega}{2} \int\limits_{0}^{T} \left[sin(3 \omega t - \phi_1)\; + sin (\phi_1)\;\right] \nonumber \\
&= \frac{F_0 \; z_0 \; \omega T}{2} sin(\phi_0) +  F_0 \; z_0 \; \omega T sin(\phi_1) \nonumber
\end{align}
This points towards the fact that the energy transacted does not depend on a single phase in a nonlinear system, and therefore the introduction of a single \emph{phase} quantity is often not very useful. As an alternative we propose the use of the eye diagrams to relate to the energy transaction for such a nonlinear system. 

\bibliography{reference}

\end{document}